\documentclass[aps,prl,twocolumn,superscriptaddress]{revtex4-2}
\usepackage[utf8]{inputenc}
\usepackage{graphicx}% Include figure files
\usepackage{amsmath}% For math environment
\usepackage{dcolumn}% Align table columns on decimal point
\usepackage{bm}% bold math
\usepackage{xcolor}% colored writing
\usepackage{siunitx}

\begin{document}
\title{Magnon-Phonon coupling in Fe$_3$GeTe$_2$}
\author{Namrata Bansal}
\affiliation{Physikalisches Institut, Karlsruhe Institute of Technology, 76131 Karlsruhe, Germany}
\author{Qili Li}
\email{Corresponding author: qili.li@kit.edu}
\affiliation{Physikalisches Institut, Karlsruhe Institute of Technology, 76131 Karlsruhe, Germany}
\author{Paul Nufer}
\affiliation{Physikalisches Institut, Karlsruhe Institute of Technology, 76131 Karlsruhe, Germany}
%\author{Hung-Hsiang Yang}
%\affiliation{Physikalisches Institut, Karlsruhe Institute of Technology, 76131 Karlsruhe, Germany}
\author{Lichuan Zhang}
\affiliation{School of Physics and Electronic Engineering, Jiangsu University, Zhenjiang, Jiangsu 212013, China}
\affiliation{Peter Gr\"unberg Institut (PGI-1) and Institute for Advanced Simulation (IAS-1) Forschungszentrum J\"ulich GmbH, D-52425 J\"ulich}
%\author{Dongwook Go}
%\affiliation{Peter Gr\"unberg Institut (PGI-1) and Institute for Advanced Simulation (IAS-1) Forschungszentrum J\"ulich GmbH, D-52425 J\"ulich}
\author{Amir-Abbas Haghighirad}
\affiliation{Institute for Quantum Materials and Technologies, Karlsruhe Institute of Technology, 76131 Karlsruhe, Germany}
\author{Yuriy Mokrousov}
\affiliation{Peter Gr\"unberg Institut (PGI-1) and Institute for Advanced Simulation (IAS-1) Forschungszentrum J\"ulich GmbH, D-52425 J\"ulich}
\affiliation{Institute of Physics, Johannes Gutenberg-University Mainz, 55099 Mainz, Germany}
\author{Wulf Wulfhekel}
\affiliation{Physikalisches Institut, Karlsruhe Institute of Technology, 76131 Karlsruhe, Germany}
\affiliation{Institute for Quantum Materials and Technologies, Karlsruhe Institute of Technology, 76131 Karlsruhe, Germany}
\date{\today}

\begin{abstract}
We study the dynamic coupling of magnons and phonons in single crystals of Fe$_3$GeTe$_2$ (FGT) using inelastic scanning tunneling spectroscopy (ISTS) with an ultra-low temperature scanning tunneling microscope. Inelastic scattering of hot carriers off phonons or magnons has been widely studied using ISTS, and we use it to demonstrate strong magnon-phonon coupling in FGT.  %Essentially, the second %derivative of the tunneling %current with respect to the %bias voltage is %proportional to the density %of states of phonons and %magnons. In the low energy %range, magnons display a %parabolic dispersion with a %smooth density of states %and phonons may display van %Hove singularities stemming %from high symmetry points in the Brillouin zone. 
We show a strong interaction between magnons and acoustic phonons %via magnetoelastic coupling 
which leads to formation of van Hove singularities originating in avoided level crossings and hybridization between the magnonic and phononic bands in this material. We identify these additional hybridization points in experiments and compare their energy with density functional theory calculations. Our findings provide a platform for designing the properties of dynamic magnon-phonon coupling in two-dimensional materials. 
\end{abstract}

\maketitle

%\section{Introduction}

%Electric resistivity in %metals is caused by the %inelastic scattering of hot %carriers with bosonic %degrees of freedom of the %crystal, i.e. generally %with phonons and in %magnetic materials %additionally with magnons. %In a conventional %description of these %scattering processes, the %dispersion and bosonic %density of states of the %two are treated separately. %While this is usually a %good approximation, the %situation changes when %large magnetocrystalline %anisotropies couple the %spin and lattice degree of %freedom such that %when the %dispersions of the phonons %and magnons match, 
%they hybridize.% and an %avoided level crossing of %the two branches arises. %This necessarily comes with %a local flattening of the %dispersion at the crossing %point and van Hove %singularities at points of %low symmetry in the %Brillouin zone.

%To resolve this coupling, high-resolution inelastic neutron scattering \cite{calder2019magnetic} has been used, which is tedious and requires large samples. An alternative detection of the coupling can be established by inelastic scanning tunneling spectroscopy (ISTS) \cite{RN1564}, which is also suitable to study small samples and two-dimensional materials. 

The neoteric exploration of two-dimensional (2D) van der Waal (vdW) magnets has recently extended the philosophy of magnonics and spintronics into the 2D realm \cite{Park_2016,Huang2017,Gong2017,Burch2018, shabbir2018long,Gibertini2019,Gong2019, Li2019intrinsic,Wang2020prospects,Wang2022}. There, due to their 2D nature, the elementary bosonic excitations in form of magnons and phonons inherit the strong symmetry breaking of the 2D material, such as anisotropic dispersion \cite{samuelsen1971spin, Wildes1998spin,calder2019magnetic, bao2022neutron,Bai2022} and chirality \cite{costa2020nonreciprocal,Yin2021chiral}. 
In magnetic 2D materials, the large uniaxial anisotropy and magnetic properties are extremely sensitive to strain engineering \cite{zhuang2016strong,Webster2018strain,Khan2020,Wang2020strain-sensitive,Hu2020enhanced,Esteras2022magnon} owing to the local nature of the magnetic moments that interact via crystal field with the surroundings. This potentially leads to an unusually high dynamic coupling of magnetic and lattice degrees of freedom in two dimensions, with strong impact on dynamical properties of 2D magnets. However, to date, very little is known about the nature and properties of elementary magnetic and lattice excitations in 2D magnetic materials. 

This necessitates the exploration of such excitations with various techniques, among which inelastic scanning tunneling spectroscopy (ISTS) emerges as one of the most powerful ones \cite{balashov2006magnon,balashov2008inelastic, Gao2008spin,Balashov2014,spinelli2014imaging,schackert2015local,jandke2016coupling}, it allows to  probe  inelastic excitations in structurally-complex 2D materials locally \cite{Jandke2019}. By now, ISTS has developed into a well established technique to locally resolve vibrons in adsorbed molecules \cite{stipe1998single}, but also has the potential to resolve bulk phonons \cite{schackert2015local,jandke2016coupling} mostly in materials with strong electron-phonon coupling, i.e. superconductors. More recently, ISTS has also been used for the detection of magnons \cite{balashov2006magnon, balashov2008inelastic,Gao2008spin,Balashov2014,spinelli2014imaging}.

% More explaination of FGT structure and properties, here.

%introduction of magnon, selection rule and its application

Among 2D magnets \cite{Park_2016,Burch2018,shabbir2018long,Gibertini2019,Gong2019,Li2019intrinsic,Wang2020prospects,Wang2022}, Fe$_3$GeTe$_2$ (FGT) fascinates with its high ferromagnetic transition temperature of 150-220 K in bulk \cite{Deiseroth2006,chen2013magnetic,tan2018hard}, even reaching room temperature with gate tuning \cite{deng2018gate}, and metallic traits \cite{Deiseroth2006,chen2013magnetic} suitable for STM investigation. %The unit %cell of the FGT consists of %a Fe$_3$Ge layer sandwiched %between two Te layers with %weak vdW force, and it %associates with the %P6$_3$/mmc space group (a = %3.991, c = 16.33 ) %\cite{Deiseroth2006, %chen2013magnetic}. Two Fe %atoms designated as %Fe$_{\alpha}$ and %Fe$_{\beta}$ are present at %the two Wyck-off positions. %Fe$_3$Ge layer forms when %the first layer make up %entirely of Fe$_{\alpha}$ %toms, arranged in a %hexagonal net while the %other Fe$_{\beta}$ and Ge %are bonded covalently in a %layer close by. 
This prolific material serves as a playground for investigating magnetic skyrmion bubbles \cite{nguyen2018visualization, ding2019observation, Park2021neel-type,yang2022magnetic,Birch2022,li2022field}, spin spirals \cite{Meijer2020}, magnetic domains \cite{fei2018two,yang2022magnetic,trainer2022relating,Birch2022,li2022field}, large uniaxial anisotropies \cite{Deiseroth2006, chen2013magnetic, León-Brito2016, tan2018hard, wang2020modification, Li2023tuning} and various prominent magnetic properties in 2D world~\cite{You2019angular, Li2023tuning, leon2016magnetic, costa2020nonreciprocal, may2016magnetic, verchenko2015ferromagnetic}.
FGT is also predicted to show strong magnetoelastic coupling \cite{zhuang2016strong}, exhibiting flat surfaces that can be easily prepared by cleaving. This makes FGT an ideal system to study the excitations due to magnons and phonons with ISTS.

In this Letter, we report a prominent coupling between magnons and acoustic phonons in 2D ferromagnetic vdW material FGT detected by laterally resolved ISTS measurements executed with scanning tunneling microscopy (STM) under ultrahigh vacuum %(${p<3}$ $\times$ $10^{-11}$ mbar) 
and low temperatures (40 mK). Supported by density functional theory calculations, we identify the peaks in inelastic spectra of FGT as fingerprints of hybrid magnon-phonon modes, which emerge at the crossing points between the magnonic and phononic bands having qualitatively different behavior in the region of small energies. Our findings not only demonstrate that FGT and 2D magnetic materials in general present an exciting platform for exploring the physics of low-energy magnon-phonon hybrid modes which may manifest, for example, in hybrid magnon-phonon topologies, but also suggest a way towards an educated material design of dynamic magnon-phonon coupling in two dimensions.

% explain ISTS

Electric resistivity in metals is caused by the inelastic scattering of hot carriers off bosonic excitations of the crystal, i.e. off phonons and in magnetic materials additionally off magnons. In a conventional description of these scattering processes, the dispersion and bosonic density of states of the two are treated separately. While this is usually a good approximation, the situation changes when large magnetocrystalline anisotropies couple the spin and lattice degree of freedom such that %when the dispersions of the phonons and magnons match, 
they hybridize.
% and an avoided level crossing of the two branches arises. This necessarily comes with a local flattening of the dispersion at the crossing point and van Hove singularities at points of low symmetry in the Brillouin zone. 
%ISTS is a well established %technique to locally %resolve vibrons in adsorbed %molecules %\cite{stipe1998single}, but %also has the potential to %resolve bulk phonons %\cite{schackert2015local,jan%dke2016coupling} mostly in %materials with a strong %electron-phonon coupling, %i.e. superconductors. 
 Within ISTS, inelastic excitations show up most clearly in the form of peak-dip pairs in the second derivative of the tunneling current w.r.t. the bias voltage, symmetrically located in bias voltage around the Fermi level \cite{wolf2011principles}. Vibrons or phonons are created by Coulomb scattering of the hot carriers with the core electrons of the atoms and thus the cross section does not depend on the spin of the tunneling electron and is typically only of the order of few percents.
%More recently, ISTS has also been used for the detection of magnons \cite{balashov2006magnon, balashov2008inelastic,Gao2008spin,Balashov2014,spinelli2014imaging}. 
In case of magnons, the scattering cross section is significantly larger as the excitation of magnons is driven by the direct exchange interaction of the tunneling electrons with the spin-polarized sample electrons. 
To identify the nature of the excitation, besides their energy, selection rules can be used. For example, phonon excitation is mediated by the electric charge of the tunneling electron, while for magnon creation by hot carriers spin selection rules apply. The latter allows to identify magnons in ISTS using spin-polarized (SP) STM \cite{Wiesendanger1990,Kubetzka2002spin-polarized,wiesendanger2009spin}.

% explain the sample

%that gives an allusion to the existence of magnons or phonons. Due to the energy range of plasmons, their contribution from these excitations eliminates, which is far off the observed excitation. To further characterize their individuality and verify the magnetic origin or the influence of the spins on these excitations, we carry out the spin-polarised (SP)-ISTS measurements which confirm the residence of magnons in this system. To complement our results, when superimposed together, the calculation for the respective dispersion bands of magnon and phonon intersect at similar energies where we have observed low energy peaks and dips in ISTS. 

% FGT

%\section*{Results and discussion}
%Paragraph 4 Explanation of experiment/work in detail.
%Fig.1

Bulk FGT has been extensively studied with STM \cite{zhang2018emergence, nguyen2018visualization, Kong2020,yang2022magnetic,trainer2022relating} and other magnetic imaging techniques, such as magnetic force microscopy \cite{leon2016magnetic, yi2016competing,chakraborty2022magnetic}, Lorentz transmission electron microscopy \cite{chakraborty2022magnetic,Ding2022tuning,li2022field} and scanning transmission x-ray microscopy \cite{Birch2022}, showing out-of-plane magnetized domains separated by rather sharp domain walls of a width of $\approx7$ nm.
High quality FGT crystals used in our work were grown using the chemical vapor transport method and were cleaved \emph{in-situ} at a base pressure of $p<3$ $\times$ $10^{-11}$ mbar leading to atomically flat surfaces with only very few step edges separated on the micro meter scale. The STM measurements were carried out using a home-built low temperature (25 mK base temperature) dilution STM system \cite{balashov2018compact}.

\begin{figure}
    \centering
    \includegraphics[width=1\linewidth]{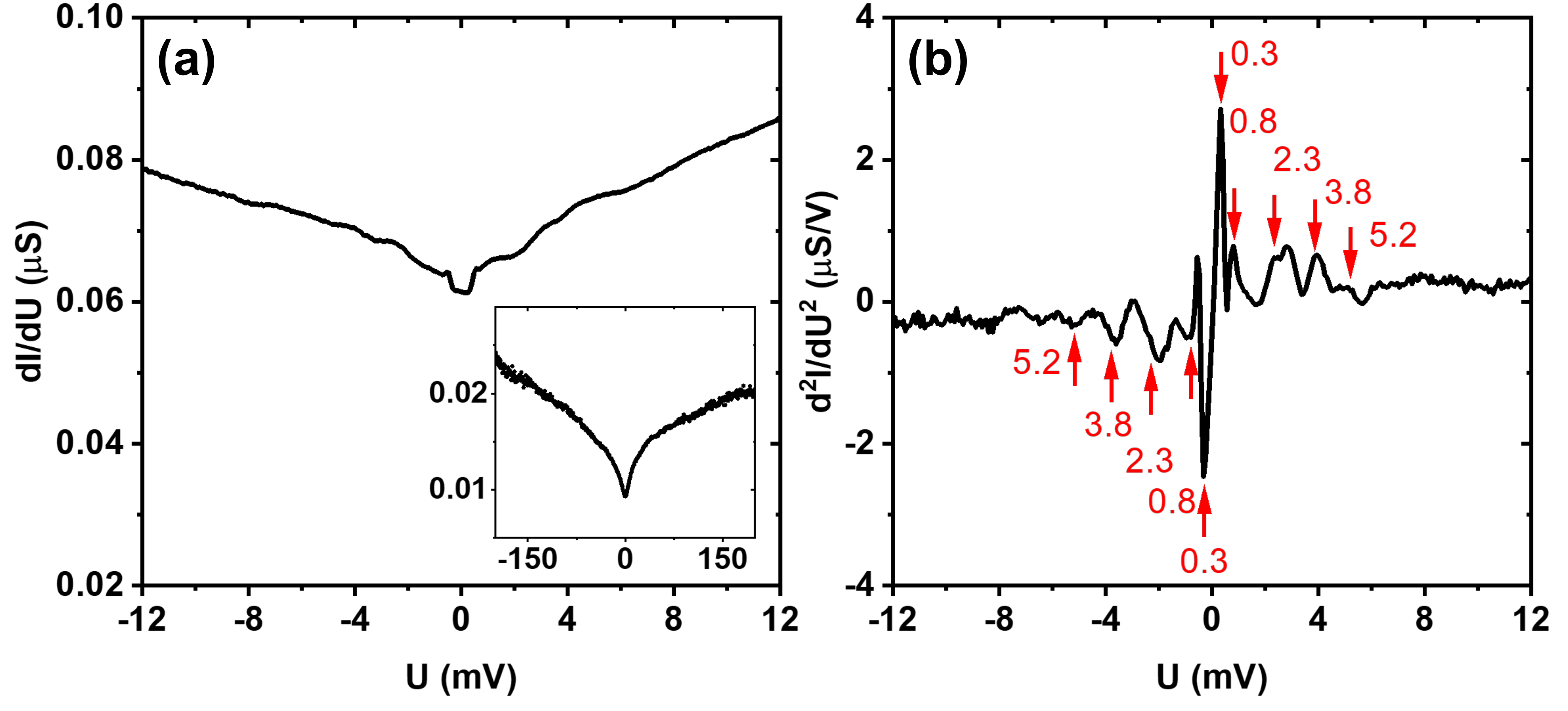}
    \caption{\textbf{Low energy excitations by inelastic electron tunneling.} (a) $dI/dU$ spectra on FGT surface. (b) $d^2I/dU^2$ spectrum recorded simultaneously. The samples were cleaved at room temperature in ultra high vaccum. The spectra were measured with W tip at 40 mK (feedback conditions $U$ = 20 \si{mV}, $I$ = 2 \si{nA}, lock-in modulation 0.5 \si{mV} at 3.421 \si{kHz}, and inset figure in (a) $U$ = 50 \si{mV}, $I$ = 1 \si{nA} and modulation amplitude 3 \si{mV}.}
    \label{fig:magnon with W tip}
\end{figure}

As in previous studies, we find a broad dip in the differential conductivity $dI/dU$ near the Fermi energy ($E_F$) \cite{zhang2018emergence, yang2022magnetic, trainer2022relating}, as depicted in the inset of Fig. \ref{fig:magnon with W tip}(a). The dip has been interpreted by Zhang {\it et al.} as a Kondo resonance \cite{zhang2018emergence}, which is disputed in the literature \cite{yang2022magnetic, trainer2022relating}, as the sample is a ferromagnet with relatively large exchange splitting of the electronic bands \cite{zhuang2016strong}. For the Kondo effect to appear, the system must obey time reversal symmetry, which is strongly broken in the ferromagnetic ground state. Moreover, the exchange interaction of localized spins should split the Kondo resonance by an energy of the order of the exchange interaction. The Kondo effect as the source of the dip can thus be safely excluded. A second explanation for the dip in $dI/dU$ could be a dip in the electronic density of states (DOS). DFT calculations for the DOS of FGT, however, exclude this \cite{yang2022magnetic}. A third possibility for the dip is inelastic tunneling, that we investigate in the following.

In elastic tunneling, the differential conductance $dI/dU$ is proportional to the DOS of the electrodes \cite{tersoff1985theory}. When variations in DOS are negligible on a small energy range around $E_F$, $dI/dU$ is constant and $d^2I/dU^2$ vanishes. An additional tunneling channel opens when the kinetic energy of the tunneling electrons is sufficient to cause an inelastic excitation. As the two channels do not interfere, an increment in the current is observed for both signs of the bias voltage. Consequently, steps appear in $dI/dU$ at the brink of the excitation energy $E_g=|eU|$, or equivalently peaks and dips appear in $d^2I/dU^2$ at $|eU|=E_g$. Fig. \ref{fig:magnon with W tip}(a) shows the differential conductivity in a small interval around $E_F$ recorded at 40 mK with a small lock-in modulation voltage of 0.5 mV. Steps at various energies can be resolved. More clearly, the $d^2I/dU^2$ spectra, recorded simultaneously with $dI/dU$ spectra, show three peaks and dips symmetrically positioned with respect to $E_F$, cf. Fig. \ref{fig:magnon with W tip}(b). Hence, the data in Fig. \ref{fig:magnon with W tip}(a) and Fig. \ref{fig:magnon with W tip}(b) are a clear evidence for the occurrence of inelastic tunneling events.

In this low energy scale of the peaks at 0.3, 0.8, 3.8 and 5.2 meV a double peak at 2.3 (see red arrows), inelastic excitation are either due to phonons or magnons, while plasmons, with eV energy, can be excluded. Note, however, that the individual pairs of peaks and dips are not identical in intensity. Typically, vibron or phonon excitations are rather symmetric in intensity. For magnons, strong asymmetries have been reported due to the spin selection rules \cite{balashov2008inelastic}. To create a magnon by inelastic exchange scattering, either a minority electron needs to be injected into the ferromagnet or a majority electron needs to be removed. Thus, the asymmetry in intensity of magnon excitation peaks relates to the spin polarization of the magnetic sample, in case an unpolarized tip is used. The calculated asymmetry, i.e. the difference in the peak area for negative and positive bias divided by their sum, taken from the data shown in Fig.\ref{fig:magnon with W tip}(b), is about $-0.1$, which is comparable with a DFT value for spin-polarization of the states of about $-0.36$ at the Fermi energy \cite{yang2022magnetic}. This hints at the magnetic origin of excitations in FGT. Further, note that when using a larger modulation voltage rather broad peak-dip pairs can be identified at energies around 6 meV and beyond. In this work, we intend to focus, however, on the low-energy sharp inelastic peaks and dips. 

\begin{figure}[t]
    \centering
    \includegraphics[width=1\linewidth]{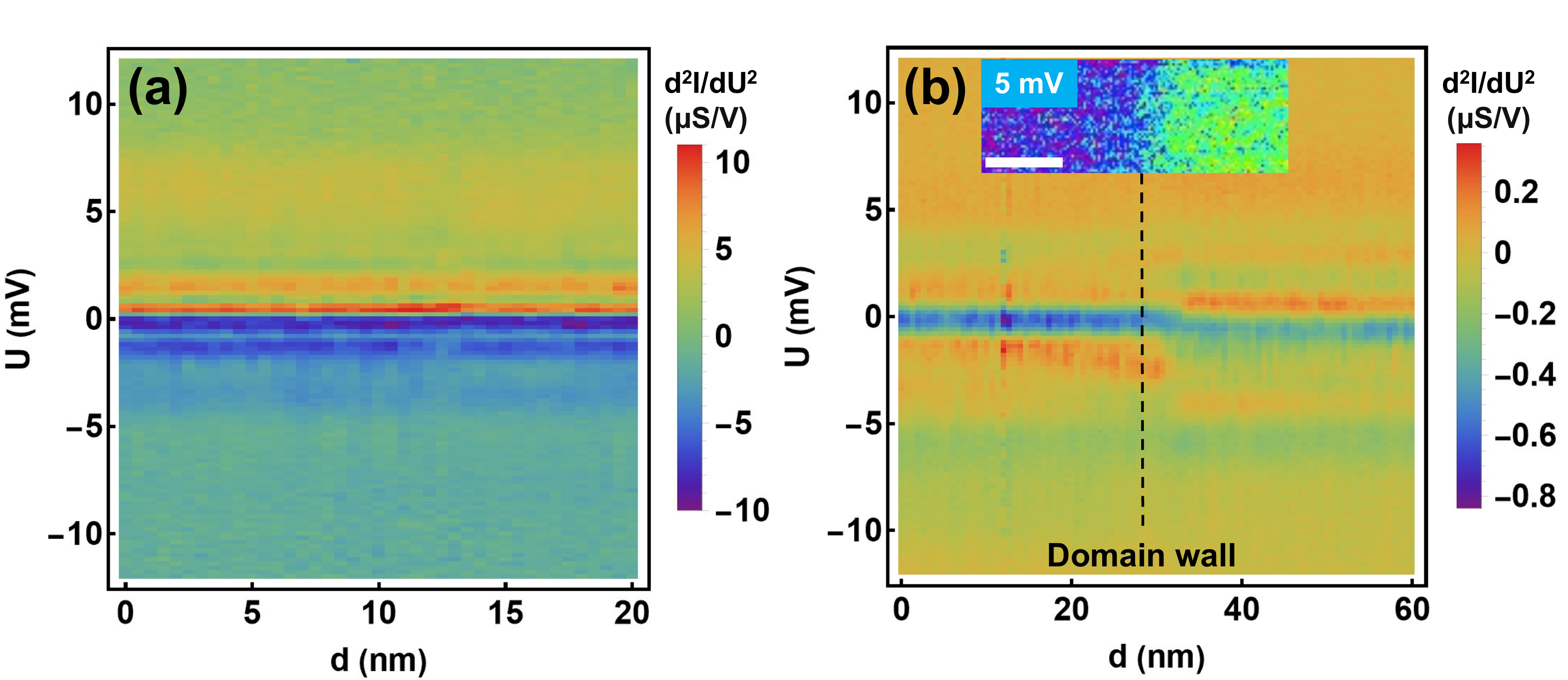}
    \caption{\textbf{Comparison between non-sp and sp results.} (a) ISTS spectra along a line (W tip,  $U$ = 20 \si{mV}, $I$ = 5 \si{nA} modulation 0.5 \si{mV} at 3.611 \si{kHz}). (b) SP-ITS spectra across a magnetic domain wall (Cr-coated tip, $U$ = 20 \si{mV}, $I$ = 0.3 \si{nA}, and modulation 1 \si{mV}). The inset figure shows $d^2I/dU^2$ mapping at 5 \si{mV} with tunneling condition $U$ = 5 \si{mV}, $I$ = 0.2 \si{nA}, and the modulation 2 \si{mV}. The white bar scale is 20 nm.}
    \label{fig:comparisons between non-sp and sp results}
\end{figure}

To further confirm the magnetic origin of these excitations, ISTS using spin-polarized STM was employed. In this technique, the STM tip is spin-polarized, such that the tunneling magneto resistance effect (TMR) \cite{julliere1975tunneling} arises, providing spin-resolved information on the sample. To polarize the tip, $\approx50$ ML of Cr were deposited on freshly prepared W tips, followed by gentle annealing \cite{Kubetzka2002spin-polarized}. The use of antiferromagnetic tips minimizes the magnetic stray-field which exert a negligible influence on the magnetization of the FGT sample \cite{yang2022magnetic}.

In the following, we compare ISTS spectra recorded along lines using unpolarized W tips and Cr coated tips. In Fig. \ref{fig:comparisons between non-sp and sp results}(a) we show the position-independent inelastic tunneling signal of the structure similar to Fig. \ref{fig:magnon with W tip}, as expected for a homogeneous sample and a spin-integrating measurement. In contrast, Fig. \ref{fig:comparisons between non-sp and sp results}(b), recorded with a spin-polarized tip, clearly shows a lateral variation in the inelastic signal corresponding to a transition of the width of about 7 nm. The inset shows a lateral map of this transition for 5 mV bias voltage. This transition was previously identified as a magnetic domain wall in SP-STM scans (not shown). The 7 nm width also agrees with previous measurements of the magnetic domain wall width \cite{yang2022magnetic}.
At the peak and dip energies seen in the inelastic spectra, we observe an intensity change at the domain wall due to spin-polarized tunneling in agreement with the spin selection rules \cite{balashov2008inelastic}. 

\begin{figure}[htb!]
    \centering
    \includegraphics[width=1\linewidth]{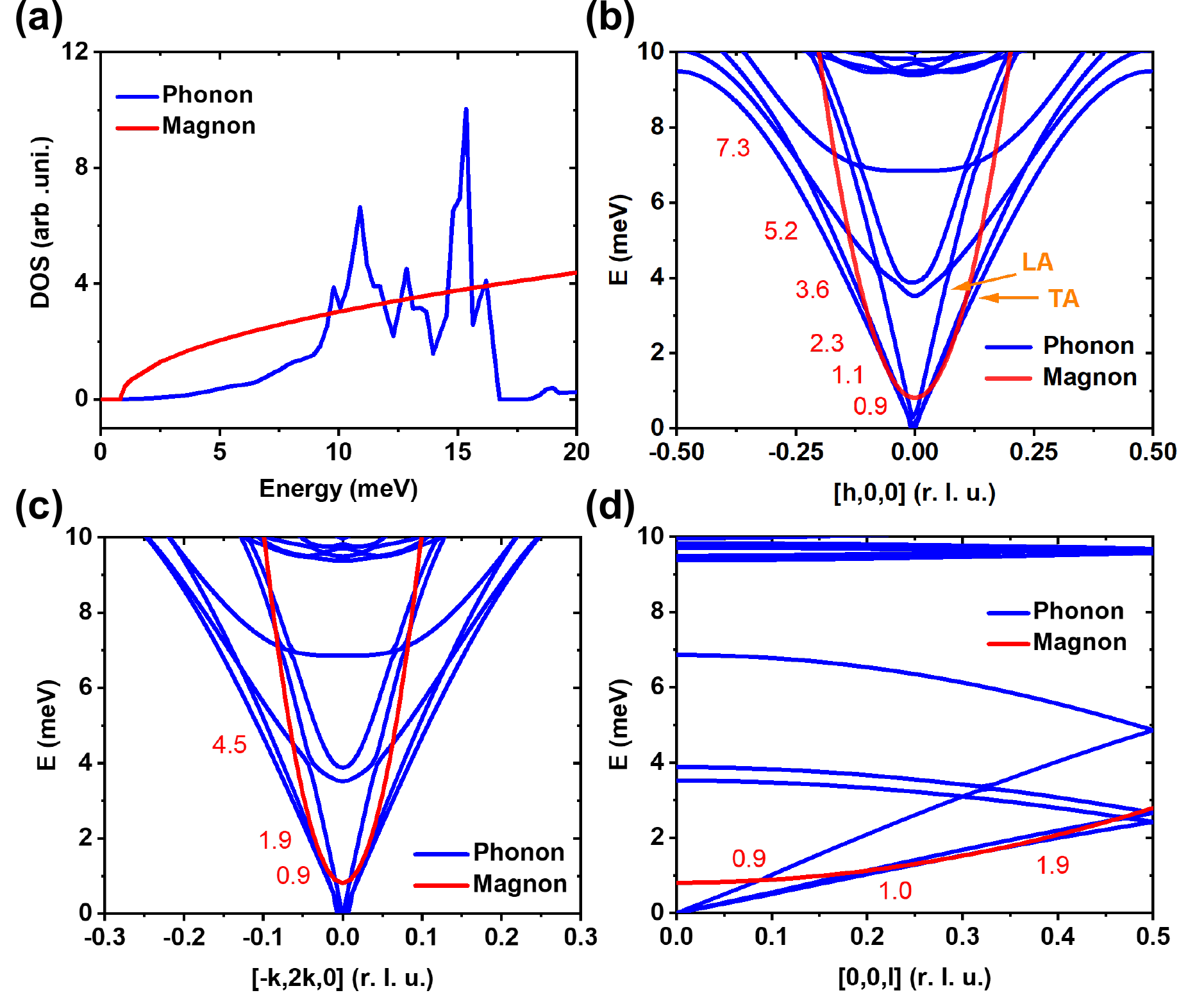}
    \caption{\textbf{Magnon and phonon band crossings in FGT.} (a) Magnon and phonon DOS in FGT. The phonon DOS is calculated from DFT. The magnon DOS is calculated from three-dimensional parabolic dispersions around the Brillouin zone center. (b) Band crossings between magnon and phonon bands along $[100]$. The spin-wave stiffness is taken at 69.0 meV\AA$^2$ from Ref. \cite{bao2022neutron}. (c) Crossings between magnon and phonon bands along $[\overline{1}20]$. The spin-wave stiffness is taken as 56.7 meV\AA$^2$ from Ref. \cite{bao2022neutron}. (d) Crossings between magnon and phonon bands along $[001]$. The spin-wave stiffness is taken as 53.6 meV\AA$^2$ from Ref. \cite{bao2022neutron}. In (b), (c) and (d), the magnon gap 0.81 \si{meV} is used to show the crossings.}
    \label{fig:Magnon and Phonon and crossings}
\end{figure}

This identifies the peaks as magnetic excitations or magnons but does not provide an insight into their energetics. Note that the parabolic dispersion of magnons around the zone center does not produce a peaked DOS, and magnons usually result in a rather smooth ISTS signal around the Fermi energy. The calculation of the DOS based on the three-dimensional parabolic dispersions around the Brillouin zone center predicts a DOS proportional to the square root of magnon energy without any peaks in the low-energy range, see Fig. \ref{fig:Magnon and Phonon and crossings}(a). 

To treat the effect of phonons, we compute their DOS and dispersion in FGT from {\it ab-initio} using the Vienna {\it ab initio} simulation package (VASP)~\cite{1996Efficient} within the local density approximation (LDA)~\cite{1981Self} for the exchange-correlation potential. The projector-augmented wave (PAW) method~\cite{1999From} is used to describe the interaction between the electrons and the nuclei, and 600 \si{eV} was selected for the kinetic energy cutoff of the plane wave expansion. The energy convergence threshold was chosen as $10^{-9}$ eV and the Brillouin zone (BZ) was sampled with a 15$\times$15$\times$5 $\Gamma$-centered Monkhorst-Pack grid~\cite{monkhorst1976special}. The shape and volume for each cell were fully optimized and the maximum force on each atom was less than 0.001eV/\AA. The phonon dispersion was calculated with the code PHONOPY~\cite{Parlinski1997first,togo2008first} in a 3$\times$3$\times$2 supercell with finite displacement method. 

Phonons might lead to sharp van Hove singularities in the meV range, as ab-initio calculations of the phonon DOS, shown in Fig. \ref{fig:Magnon and Phonon and crossings}(a) in blue, indicate. There are peaks between 10 and 15 \si{meV}, which are consistent with the results of inelastic neutron scattering \cite{bao2022neutron,Bai2022}. Note that we did not observe these phonon peaks in the ISTS shown in Fig. \ref{fig:magnon with W tip}. This implies a weak electron-phonon coupling. Below 5 meV the phonon DOS is smooth as the lowest energy bands for transversal and longitudinal acoustic phonons disperse linearly in that energy range. When allowing for magnon-phonon coupling, however, the situation drastically changes. In case of a crossing between the phonon and magnon bands,  magnetoelastic coupling hybridizes the two distinct modes and an avoided level crossing is expected. At these so called ``hot spots" in the Brillouin zone, magnetization strongly couples to the lattice, which is also responsible for parts of Gilbert damping \cite{McMichael2002}. 

We predict that in FGT such avoided level crossings with diabolic points cause a high DOS due to hybrid excitaions. In Fig. \ref{fig:Magnon and Phonon and crossings}(b) we plot the {\it ab-initio} phonon dispersion of FGT together with an experimental dispersion curve of the magnons obtained from neutron scattering \cite{bao2022neutron}. The measured spin-wave stiffness from neutron scattering is about 69.0 meV\AA$^2$ along the $[100]$-direction, while the spin-wave gap calculated from the uniaxial anisotropy of the sample with $K_u=1.46\times 10^6$ \si{J/m^3} and the saturation magnetisation $M_s=3.9\times 10^5$ \si{A/m} \cite{León-Brito2016} using Kittel's formula $\omega=\gamma(2 K_u/M_s - \mu_0 M_s)$ % please use SI units 
\cite{Beaujour2009,Zakeri2014} amounts to 0.81 \si{meV}, where $\gamma=g\mu_B/\hbar$ is the gyromagnetic ratio. $K_u$ is the uniaxial anisotropy. Thus, the plot contains no fitting parameters. The plot predicts band crossings of the acoustic phonons -- with their linear dispersion -- with the parabolic but gapped magnon spectrum very close to the $\Gamma$-point at 0.9 meV along all directions (see Fig. \ref{fig:Magnon and Phonon and crossings}(c)) in agreement with the observed peak and dip in the inelastic spectrum. As the quadratic magnon dispersion rises eventually faster, more crossings may appear along the different directions. 
Note that in the $[001]$-direction, i.e. normal to the vdW layers, the exchange is weak so that a second crossing with all except the lowest phonon band can be excluded (see Fig. \ref{fig:Magnon and Phonon and crossings}(d)). Along the high-symmetry $[100]$-direction, second crossings are expected at 2.3, 3.6 and 5.2 meV in excellent agreement with the experimental observations.
Crossing along the lower symmetry $[\overline{1}20]$-direction are expected to have a lower spectral weight and the corresponding peaks are not clearly resolved, or no hybridization occurs in these directions.
The theoretically predicted crossing points are in a very good comparison with the position of the experimental inelastic peaks and dips shown in Fig. 1, which clearly demonstrates for the first time the presence of strong magnon-phonon coupling in one of the most prolific 2D materials. Note that the lowest energy peak and dip pair at 0.3 meV cannot be due to magnon-phonon coupling. Instead, it is found at similar energies in most samples speaking for a dynamical Coulomb blockade effect \cite{ schackert2015local,Ast2016sensing,Senkpiel2020dynamical}.
Interestingly, also optical phonons cross the magnon dispersion at higher energies, where also neutron scattering data shows that the magnon dispersion becomes heavily damped \cite{calder2019magnetic,bao2022neutron,Bai2022}. At these energies, no sharp van Hove singularities are expected, but magnon-phonon coupling for optical phonons has been detected using Raman spectroscopy \cite{Du2019lattice}.

In summary, strong magnon-phonon coupling has been observed with ISTS for low energies near the Zone center at crossings of the dispersion of acoustic phonons and magnons. Spin-polarized measurements confirm the magnetic nature of these hybrid excitations and domain walls can also be resolved in the inelastic signal. As ISTS is a local technique, it offers the possibility to study confined modes in structured samples. One can imagine that strain engineering allows to localize magnons in bulk or even single layer FGT in form of magnon wave guides or inversely, to use magnetic domain walls for phonon localization. Finally, the hybrid bosonic excitations can lead to a chiral nature of the phonons part with the possibility to guide transport of angular momentum of the lattice.  
%In principle, localized magnons can also be present inside the narrow magnetic domain wall, as has been discussed in the literature \cite{niez1976localized}. ISTS can, in principle, detect these modes and they would show up in the spectra as additional peak-dip pairs localized at the domain wall. {\bf May be one concluding sentence w.r.t. to the relevance to the future?} Our results shed light on the potential application for acoustic magnonics in FGT.

This work was funded by the Deutsche Forschungsgemeinschaft (DFG) in the framework of TRR 288-422213477 (Project B06 and B03), CRC 1238-277146847 (Project C01) and the Cluster of Excellence Matter and Light for Quantum Computing (ML4Q) EXC 2004/1-390534769.

\bibliography{main}

\end{document}